\documentclass[epj]{svjour}

\usepackage{graphics}

\def\ET{\mbox{$E_T$}}
\def\pT{\mbox{$p_T$}}

\def\qhat{\mbox{$\hat{q}$}}
\def\qTsq{\mbox{$\langle{q_T^2}\rangle$}}

\def\sqrtsNN{\mbox{$\sqrt{s_\mathrm{_{NN}}}$}}
\def\sqrts{\mbox{$\sqrt{s}$}}

\def\Nbinary{\mbox{$N_{bin}$}}
\def\NbinaryMean{\mbox{$\langle\Nbinary\rangle$}}

\def\pizero{\mbox{$\pi^0$}}
\def\kzeros{\mbox{$K^0_s$}}

\def\pp{\mbox{$p+p$}}

\def\dNchdeta{\mbox{$dN_{ch}/d\eta$}}

\def\vtwo{\mbox{$v_2$}}

\def\RAA{\mbox{$R_{AA}(\pT)$}}

\def\RCP{\mbox{$R_{CP}(\pT)$}}

\def\lt{\mbox{$<$}}
\def\gt{\mbox{$>$}}

\begin{document}

\title{Jets in Nuclear Collisions: Status and Perspective
\thanks{Summary talk on jet physics, Hard Probes 2004, Ericeira, Portugal, Nov. 4-10 2004}}

\author{Peter Jacobs}

\institute{CERN and Lawrence Berkeley National Laboratory, \email{pmjacobs@lbl.gov}}

\date{Received: date / Revised version: date}

\abstract{I review the status and future directions of jet-related measurements in high energy nuclear collisions 
and their application as a probe of QCD
matter.  
}

%{13.87.-a}{Jets in large-Q2 scattering} 
%{25.75.Nq}{Quark deconfinement, quark-gluon plasma production, and phase transitions}

\maketitle

%--------------------------------------------------------

\section{Introduction}
\label{intro}

Long before the startup of RHIC, jet physics was widely expected to
play an important role in the study of QCD matter at collider
energies. It is nevertheless surprising just how crucial a role that
has turned out to be. Strong modification of jet fragmentation in high
energy nuclear collisions is now well-established. The magnitude of
the effects and the statistical abundance of jet-related observables
at RHIC have allowed characterization of this modification in
significant detail, indicating that it arises mainly from interactions
of high energy partons with the medium prior to
hadronization. Perturbative QCD calculations incorporating partonic
energy loss via medium-induced gluon radiation (``jet quenching'') are
able to reproduce the broad features of the measurements, under the
condition that the early hot phase of the collision fireball has gluon
density $\sim$30-50 times that of cold nuclear matter. The jet
quenching phenomena taken together with other measurements at RHIC, in
particular evidence for early equilibration and near-ideal
hydrodynamic flow, suggest that dense, locally equilibrated matter
dominated by partonic degrees of freedom is created in high energy
nuclear collisions
\cite{Adcox:2004mh,Back:2004je,Arsene:2004fa,adams:2005dq}. 

These conclusions are at present somewhat qualitative,
however. Important aspects of partonic energy loss theory are as yet
untested. In particular, the induced radiation and its interaction with
the medium have not been clearly identified, and the expected
variation of energy loss with partonic species (gluon, light quark,
heavy quark) is only beginning to be explored. Significant
medium-induced effects in the ``intermediate
\pT'' regime at RHIC (\pT$\sim2-5$ GeV/c) have not yet been fully 
understood, though they appear to arise from the interaction between
hard and soft processes and provide a new window into mechanisms that
could drive the system to equilibrium.

This conference is therefore timely. The qualitative discovery period
for jet physics in nuclear collisions at RHIC is perhaps winding down
and the most significant progress may now come from quantitative
comparison of measurement and theory. I will summarize some of the
main experimental results to date, highlighting recent developments
presented at this conference, and briefly look ahead to the RHIC II
and LHC eras. The field of partonic interactions in matter is large
and growing and there are important areas I will not cover, for
instance forward physics which is discussed in \cite{BlandHardProbes}.

%------------------------------------------------------------------------------
\section{Hadron Suppression and Partonic Energy Loss}
\label{Inclusive}

The inclusive \pT\ spectrum in a high energy hadronic collision is
dominated by hadrons carrying a large fraction of the energy of their
parent partons ($\langle{z}\rangle\sim0.7$ for the inclusive hadron
spectrum compared to $\langle{z}\rangle\sim0.3$ for the highest energy
hadron in a light quark jet, where
$z=p_{hadron}/E_{jet}$). Modification of the partonic spectrum due to
energy loss in matter is therefore reflected directly in the
suppression of inclusive hadron yields at high \pT. Strong
medium-induced suppression (factor $\sim5$) has been measured in
nuclear collisions at RHIC, providing the primary evidence for
partonic energy loss
\cite{Adcox:2004mh,Back:2004je,Arsene:2004fa,adams:2005dq}. In contrast,
prompt photon production has been observed not to be suppressed in
nuclear collisions \cite{Frantz:2004gg}, consistent with the
interpretation of high \pT\ hadron suppression resulting from the
interaction of colored partons in the medium.

\begin{figure}
\resizebox{0.5\textwidth}{!}{\includegraphics{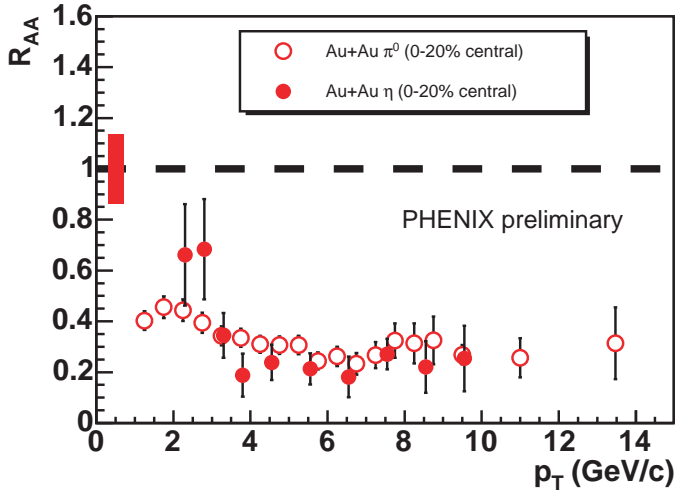}}
\caption{Binary collision-scaled ratio \RAA\ of \pizero\ and $\eta$ inclusive yields, for central Au+Au 
relative to \pp\ collisions \cite{BueschingHardProbes}.}
\label{fig:PHENIXeta}
\end{figure}

\begin{figure*}
\resizebox{0.5\textwidth}{!}{\includegraphics{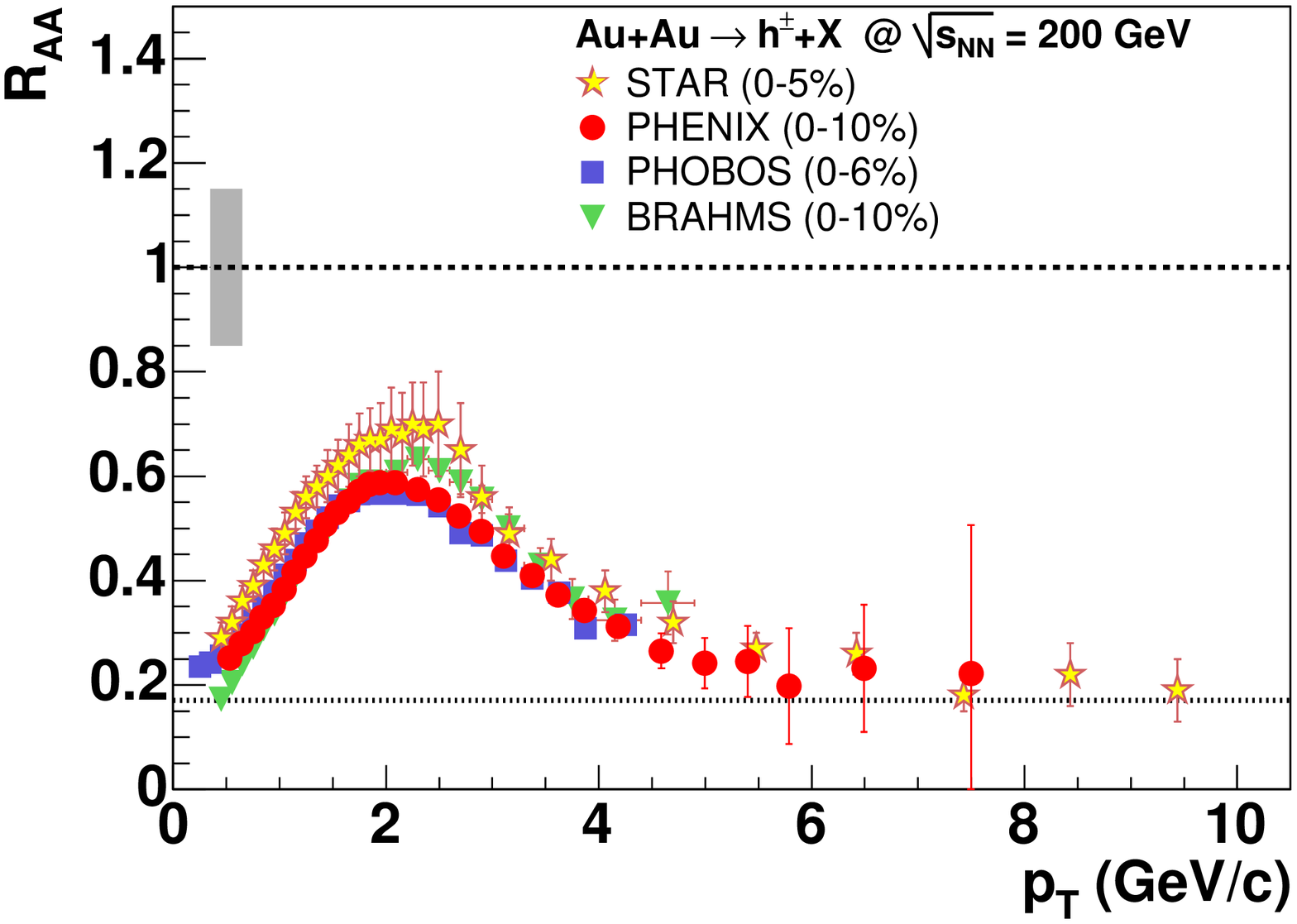}}
\resizebox{0.5\textwidth}{!}{\includegraphics{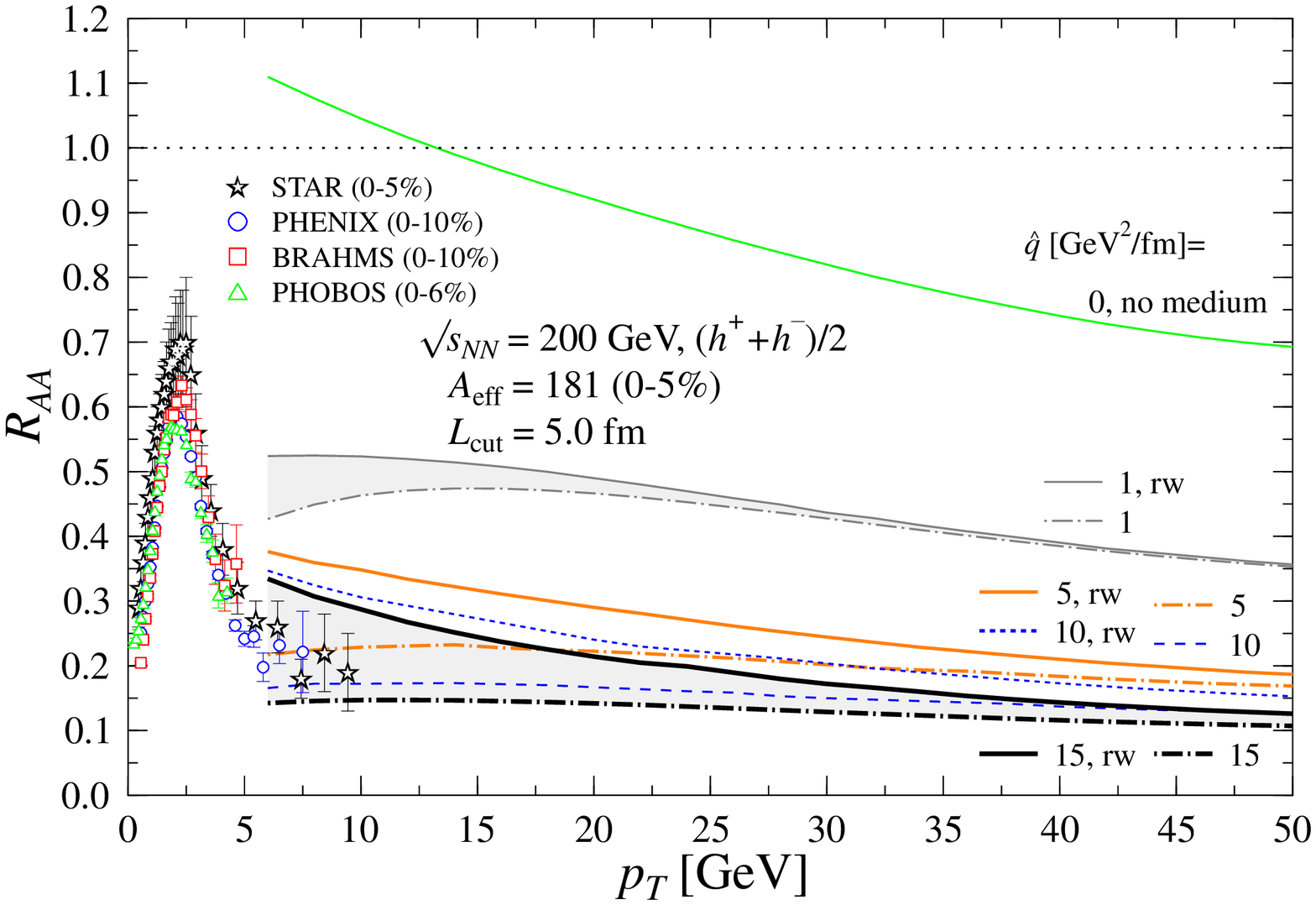}}
\caption{Left: \RAA\ in central Au+Au collisions for charged hadrons \cite{d'Enterria:2004fm}. 
Right: the same data compared to a BDMPS-based calculation with varying transport
coefficient \qhat\ \cite{Eskola:2004cr}. The pair of curves for each
\qhat\ indicates the theoretical uncertainty due to finite energy
corrections (``rw'' refers to reweighting the gluon emission
probability distribution so that the total radiated energy does not
exceed the energy of the parent parton).}
\label{fig:RAACharged}
\end{figure*}

Figure \ref{fig:PHENIXeta} shows recent measurements of hadron
suppression in nuclear collisions from PHENIX
\cite{BueschingHardProbes}. \RAA\ is the ratio of the inclusive yield
in nuclear collisions to the yield in p+p collisions, scaled by the
number of binary collisons \NbinaryMean\ to remove uninteresting
geometric effects. \RAA=1 correpsonds to no nuclear modification. The
well-known factor $\sim5$ suppression of
\pizero\ now extends to \pT=14 GeV/c, showing no significant
\pT-depdendence. New at this conference is \RAA\ for $\eta$ mesons,
whose suppression is in quantitative agreement with that for
\pizero. The $\eta/\pizero$ yield ratio is independent of collision 
system \cite{BueschingHardProbes}, consistent with hadron production
being dominated by jet fragmentation in vacuum even for central Au+Au
collisions. Figure \ref{fig:RAACharged}, left panel, shows the equally
well-known \RAA\ for charged hadrons in central Au+Au collisions,
whose suppression for \pT\gt5 GeV/c is in quantitative agreement with
that for \pizero. The growth toward lower \pT\ for charged hadron
\RAA\ arises from different suppression for protons and
pions in this region; more on this point below.

The inclusive \pizero\ spectrum in 200 GeV p+p collisions is well
represented by NLO pQCD calculations incorporating suitable
fragmentation functions \cite{Adler:2003pb}, meaning that the
reference spectrum for the inclusive suppression measurements is well
understood. Medium-induced radiative energy loss has been incorporated
into factorized pQCD calculations in two approximations, multiple soft
interactions (BDMPS) and few hard scatterings (GLV opacity expansion),
and the modification of the fragmentation function measured in DIS
scattering off nuclei has been related to its modification in heavy
ion collisions via a twist expansion (Wang and Wang) (for recent
reviews see
\cite{Baier:2000mf,Gyulassy:2003mc,Kovner:2003zj}). Within the BDMPS
framework the medium is characterized by a transport coefficient
$\qhat=\qTsq/\lambda$, where $\qTsq$ is the mean squared transverse
momentum generated by interactions in the medium over mean free path
$\lambda$. \qhat\ is related to the energy density via
$\qhat=c\epsilon^{3/4}$, where $c\sim2$ for an ideal pion gas and
weakly interacting quark-gluon plasma \cite{Baier:2002tc} but takes
values $c\sim10-20$ from estimates based on hydrodynamics
\cite{Eskola:2004cr}. In the high energy limit for a static medium of spatial dimension $L$
the average radiative energy loss is
$\Delta{E}\sim\alpha_s\qhat{L^2}/2$
\cite{Baier:2002tc}, though this expression is only qualitative since
finite kinematic limits and dynamic expansion effects generate
significant modifications to the radiated spectrum and reduce the $L$
dependence to linear. However, the induced radiation spectrum for a
dynamically expanding medium is the same as that for a static medium
having \qhat\ equal to the time-averaged \qhat\ for the dynamic
case\cite{Salgado:2003gb}.

With suitable choice of model parameters, the multiple soft scattering
and opacity expansion approximations generate similar effects for
energy loss and jet broadening
\cite{Salgado:2003gb}, and calculations applying either approach are successful in describing the 
inclusive hadron suppression and its systematic dependence on
centrality and collision energy
\cite{WangHardProbes,WiedemannHardProbes}. Figure \ref{fig:RAACharged}, 
right panel, shows \RAA\ calculated in the
BDMPS framework compared to the data in the left panel
\cite{Eskola:2004cr}. For $\pT\gt\sim5$ GeV/c the level of suppression
is described equally well by all choices $\qhat\gt5$ GeV$^2$/fm, due
to the complete dissipation of the energy of hard-scattered partons in
the core of the fireball for large \qhat\ \cite{Eskola:2004cr}. In
this and similar calculations the observed hadron population at high
\pT\ is biased towards fragments of partons scattered near the
surface of the fireball and directed outward, thereby suffering
relatively little energy loss (see also
\cite{Adler:2002tq,Drees:2003zh,Dainese:2004te}). 

Inclusive hadron production is therefore subject not only to
fragmentation bias (prefering quark to gluon jets) but also to geometric
bias towards peripheral production and small energy loss. Its
information content is consequently limited: for a core that is
sufficiently opaque, measurement of the leading hadron suppression has
little sensitivity to just how opaque it is \cite{Eskola:2004cr} and
study of partonic interactions with the medium must rely on additional
measurements, in particular correlations.

Figures \ref{fig:PHENIXeta} and \ref{fig:RAACharged} show that \RAA\ at
high \pT\ is largely independent if \pT. While several pQCD-based
calculations incorporating radiative energy loss are able to reproduce
the \pT-independence of the suppression, they ascribe it variously
\cite{ColeHardProbes} to the interplay between quenching, shadowing, 
and Cronin enhancement \cite{Vitev:2002pf}, absorption of thermal gluons at
moderate \pT\ \cite{Wang:2001cs}, the dominance of geometry due to
strong absorption
\cite{Dainese:2004te,Eskola:2004cr}, and the interplay between energy 
loss and steepening of the partonic spectrum due to phase space
limitations \cite{Eskola:2004cr}. It is important to disentangle these
mechanisms, for instance to isolate the effects of stimulated emission
and absorption which can drive parton thermalization
\cite{ColeHardProbes}. Variation of
\sqrtsNN\ provides a potentially signficiant test of the models:
lowering \sqrtsNN\ steepens the underlying partonic spectrum which is
expected to generate stronger Cronin enhancement and stronger hadron
suppression at fixed energy loss, while the initial gluon density of
the medium may be lower, resulting in reduced energy loss.

In 2004 RHIC carried out a two week run of Au+Au at \sqrtsNN=62.4 GeV,
chosen to match the top ISR energy where extensive p+p reference data
are available. It turns out, unfortunately, that disagreement among
the ISR datasets limits the systematic uncertainty of the p+p reference
to about 30\%
\cite{dEnterriaHardProbes,GagliardiHardProbes}. Nevertheless, strong 
suppression of \pizero\ and charged hadron inclusive yields is
observed in 62 GeV Au+Au central collisions for $\pT\gt\sim5$ GeV/c,
beyond the baryon enhancement region
\cite{Back:2004ra,GagliardiHardProbes,BueschingHardProbes}. In particular,
the binary scaled central to peripheral ratio \RCP, which does not rely
on the ISR reference, exhibits the same magnitude of suppression for 62
and 200 GeV Au+Au collisions\cite{GagliardiHardProbes}. The
measured high \pT\ suppression at 62 GeV broadly agrees with theoretical {\it
predictions}
\cite{Dainese:2004te,dEnterriaHardProbes,Vitev:2004gn,Adil:2004cn,Wang:2004yv}, 
though the limited statistical reach of the current 62 GeV dataset
does not support more quantitative conclusions. The RHIC program will
return to 62 GeV Au+Au for a longer run and additionally will acquire
its own high statistics p+p reference data at 62 GeV.

For SPS fixed target data at yet lower $\sqrtsNN\sim17$ GeV,
reconsideration of the available p+p reference data brings \pizero\
suppression into agreement with partonic energy loss calculations
incorporating Bjorken model estimates of the energy density
\cite{dEnterriaHardProbes}, though the statistics in the hard
scattering region are quite limited. Overall, precision study of the
\sqrtsNN\ dependence of inclusive hadron suppression remains open.

%------------------------------------------------------------------------------
\section{Jet-like Correlations}
\label{Correlations}

\begin{figure*}
\resizebox{0.5\textwidth}{!}{\includegraphics{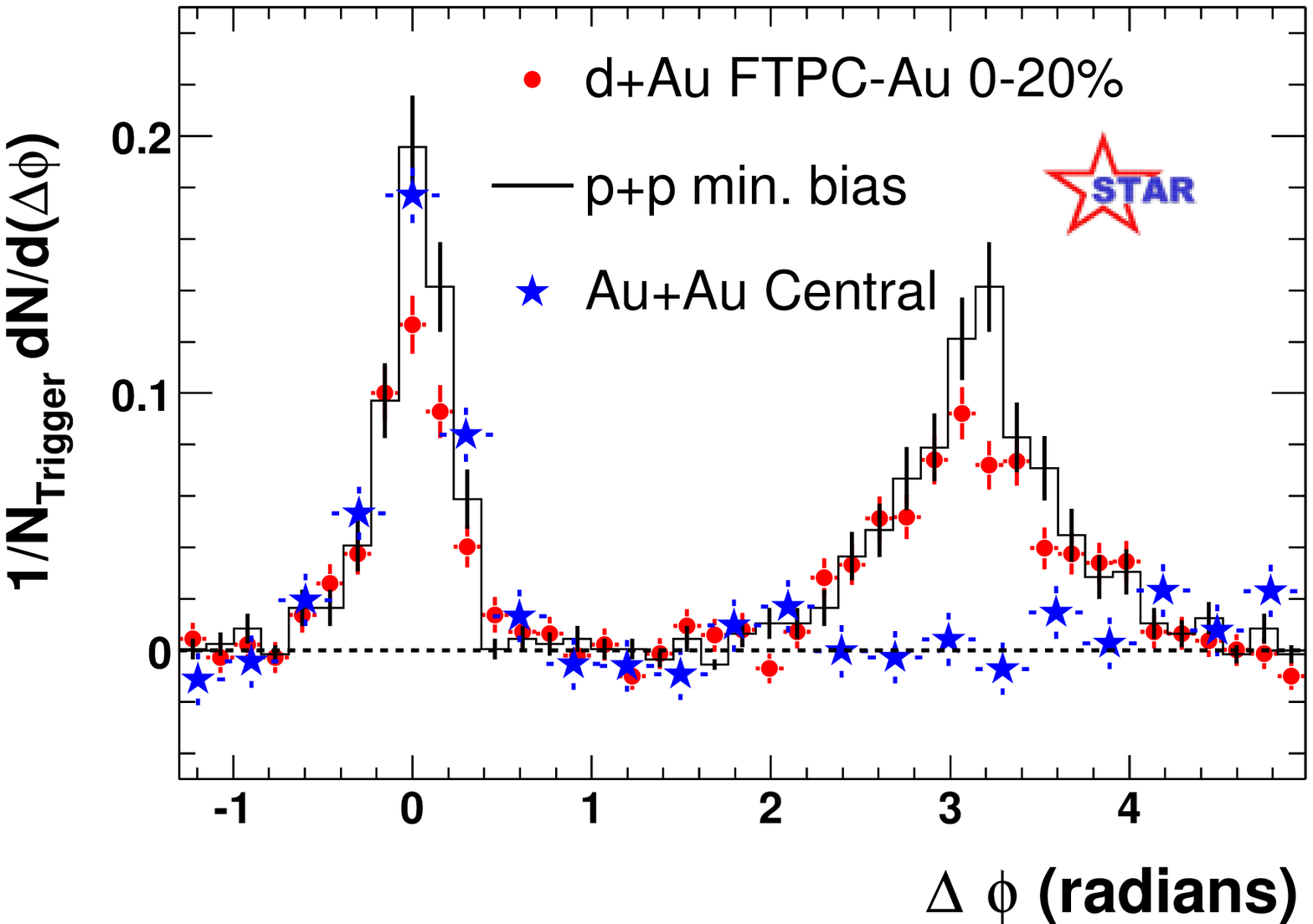}}
\resizebox{0.5\textwidth}{!}{\includegraphics{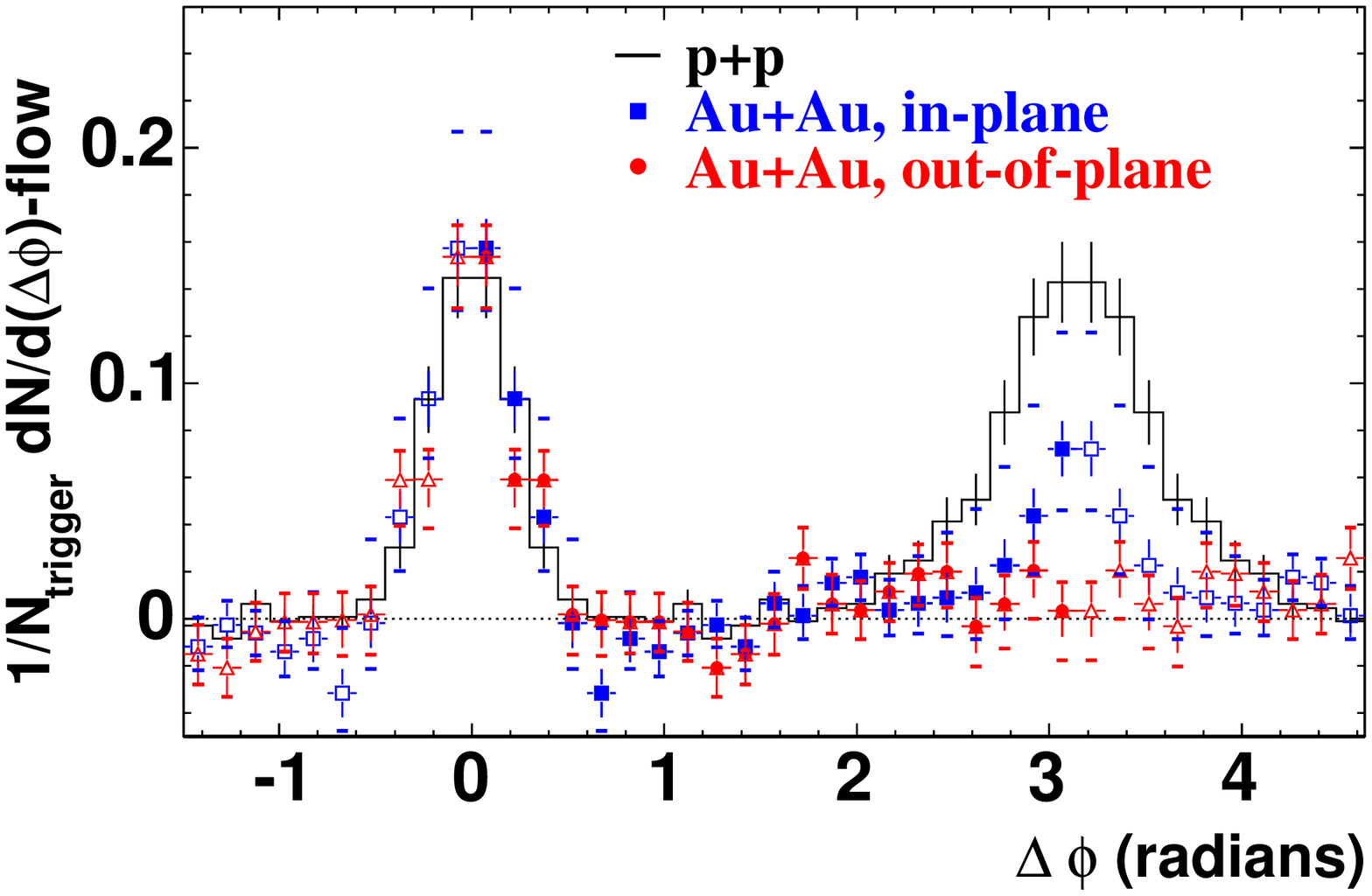}}
\caption{Relative azimuthal angle distribution for high \pT\ dihadrons at \sqrts=200 GeV. Left: p+p, 
d+Au and central Au+Au collisions \cite{Adams:2003im}. Right: trigger
particle in or out of reaction plane in non-central Au+Au collisions
\cite{Adams:2004wz}.}
\label{fig:Corr}
\end{figure*}

Hard partonic scattering generates pairs of recoiling jets which are
to first order back-to-back in azimuth. While full jet reconstruction
is challenging in high energy nuclear collisions due to the large
combinatorial background, additional insight into partonic energy loss
can be gained by studying jet-like correlations of high \pT\ pairs
(``dihadrons''). Figure \ref{fig:Corr}, left panel, shows the dihadron
distribution in relative azimuthal angle ($\Delta\phi$), with trigger
$\pT^{trig}\gt4$ GeV/c and associated hadron $2\lt\pT\lt{\pT^{trig}}$
\cite{Adams:2003im}. Dihadrons at small relative angle
$\Delta\phi\sim0$ are drawn from the same jet cone. The geometric bias
discussed above should also affect the dihadron trigger, and the approximate
similarity of the $\Delta\phi\sim0$ correlation seen for p+p, d+Au and
central Au+Au collisions is consistent with negligible partonic energy
loss and fragmentation in vacuum in the latter case.

Back-to-back dihadrons ($\Delta\phi\sim\pi$) are drawn from the
mutually recoiling jet pair. The presence of back-to-back correlations
in p+p and d+Au but their absence in central Au+Au collisions suggests
strong suppression of the leading fragments of the recoiling jet,
consistent with significant medium-induced partonic energy loss. The
geometric bias of the trigger also plays an important role in this
case: since the trigger is biased towards short path length in the
medium, the recoil is biased towards large path length through
the core of the fireball, thereby enhancing the energy loss
effects. 

The observed suppression of back-to-back dihadrons at high
\pT\ is sometimes mis-stated as the suppression of back-to-back
jets. The recoil jet has not been suppressed, rather its fragmentation
has apparently been softened. Its energy and momentum should be
conserved and should reappear at lower \pT. This has been demonstrated
qualitatively in \cite{Adams:2005ph}, where the soft recoiling hadrons
in central collisions are seen to be distributed somewhat more broadly
in azimuth and softened in \pT\ relative to similar distributions in
p+p, with the azimuthal distribution consistent with no dynamical
correlations beyond simple momentum conservation
\cite{Borghini:2000cm}. There has recently been discussion of
additional structure in the recoil azimuthal distribution at low \pT\
which may be generated, for instance, by sonic shock waves resulting
from deposition of the recoiling jet energy in an ideal, low-viscosity
fluid \cite{Casalderrey-Solana:2004qm}. Such measurements require
careful subtraction of background, however, and the significance of
the features seen in the data is under discussion.

Figure \ref{fig:Corr}, right panel, shows a similar high \pT\ dihadron
analysis for non-central Au+Au collisions, with the trigger now
constrained to lie within restricted azimuthal intervals centered on
the reaction plane orientation or orthogonal to it
\cite{Adams:2004wz}. The small angle $\Delta\phi\sim0$ correlation is
seen to be independent of the orientation of the trigger while the
recoil $\Delta\phi\sim\pi$ correlation is sensitive to it, with
stronger suppression out-of-plane. Again taking into account the
geometric trigger bias, an out-of-plane trigger corresponds to larger
in-medium path length for the recoil than an in-plane trigger. This is
the clearest indication to date of the in-medium path length
dependence of hadron suppression. Though the systematic uncertainties
in the figure are large and strongly (anti-)correlated between the
trigger selections, this is nevertheless a promising observable for
the high luminosity RHIC Run 4 analysis, where the study of higher
\pT\ correlations with correspondingly smaller background corrections
may enable a detailed measurement of the path length dependence of the
suppression.

One of the striking observations at RHIC is the large
medium-induced enhancement of the baryon/meson inclusive yield ratio
in the intermediate \pT\ region $\sim2\lt\pT\lt5$ GeV/c
\cite{Adler:2003kg,Adams:2003am}, which reverts at higher
\pT\ to the ratio observed for jet fragmentation in vacuum. Stated
differently, mesons at intermediate \pT\ exhibit strong suppression
while baryons are much less suppressed, scaling roughly as the number
of binary collisions. Measurements of several hadron species indicate
that this is a meson/baryon distinction, not dependent upon particle
mass \cite{GagliardiHardProbes,VelkovskaHardProbes}. Additionally,
elliptic flow \vtwo\ (the azimuthal modulation of the inclusive yield
relative to the reaction plane) is seen to be large at intermediate
\pT\ in non-central collisions, apparently exceeding the initial
spatial anisotropy \cite{Adams:2004wz}, with \vtwo\ for baryons larger
than that for mesons \cite{Adams:2003am}. The large \vtwo\ values
cannot be fully accounted for by so-called ``non-flow'' contributions
(e.g. intra-jet correlations at high \pT) \cite{Adams:2004wz}. It is
difficult to reconcile these observations with a scenario in which
hadron production results dominantly from jet fragmentation and its
modification due to partonic energy loss: stronger suppression of
inclusive meson than baryon distributions would correspond to larger
energy loss in medium and therefore larger azimuthal modulation \vtwo\
for mesons, in contrast to the measurements.

An alternative view of hadron production at intermediate \pT\ emerges
from the observation that \vtwo\ for various meson and baryon species
approximately follows a common distribution when rescaled as $\vtwo/n$
vs $\pT/n$
\cite{Adams:2003am,GagliardiHardProbes}, where $n=3$ for baryons and $n=2$ for mesons. Such scaling is 
expected from models in which hadronization
occurs via coalescence or recombination of constituent quarks
\cite{Voloshin:2002wa,Fries:2003vb,Hwa:2004ng,Greco:2003xt}. 
The models consider recombination of dressed quarks from two origins:
a collectively flowing thermal medium generating an exponential \pT\
distribution, and fragmentation of hard scattered partons generating a
power law \pT\ distribution. Because baryons carry the momentum of
three recombining quarks compared to two for mesons, generically in
these models the thermal component is dominant to higher hadronic \pT\
for baryons than mesons. The enhancement at intermediate \pT\ of the
baryon/meson inclusive yield ratio, the larger baryon
\vtwo, and the constituent quark number scaling of \vtwo\ emerge
naturally from the recombination approach.

\begin{figure*}
\resizebox{0.5\textwidth}{!}{\includegraphics{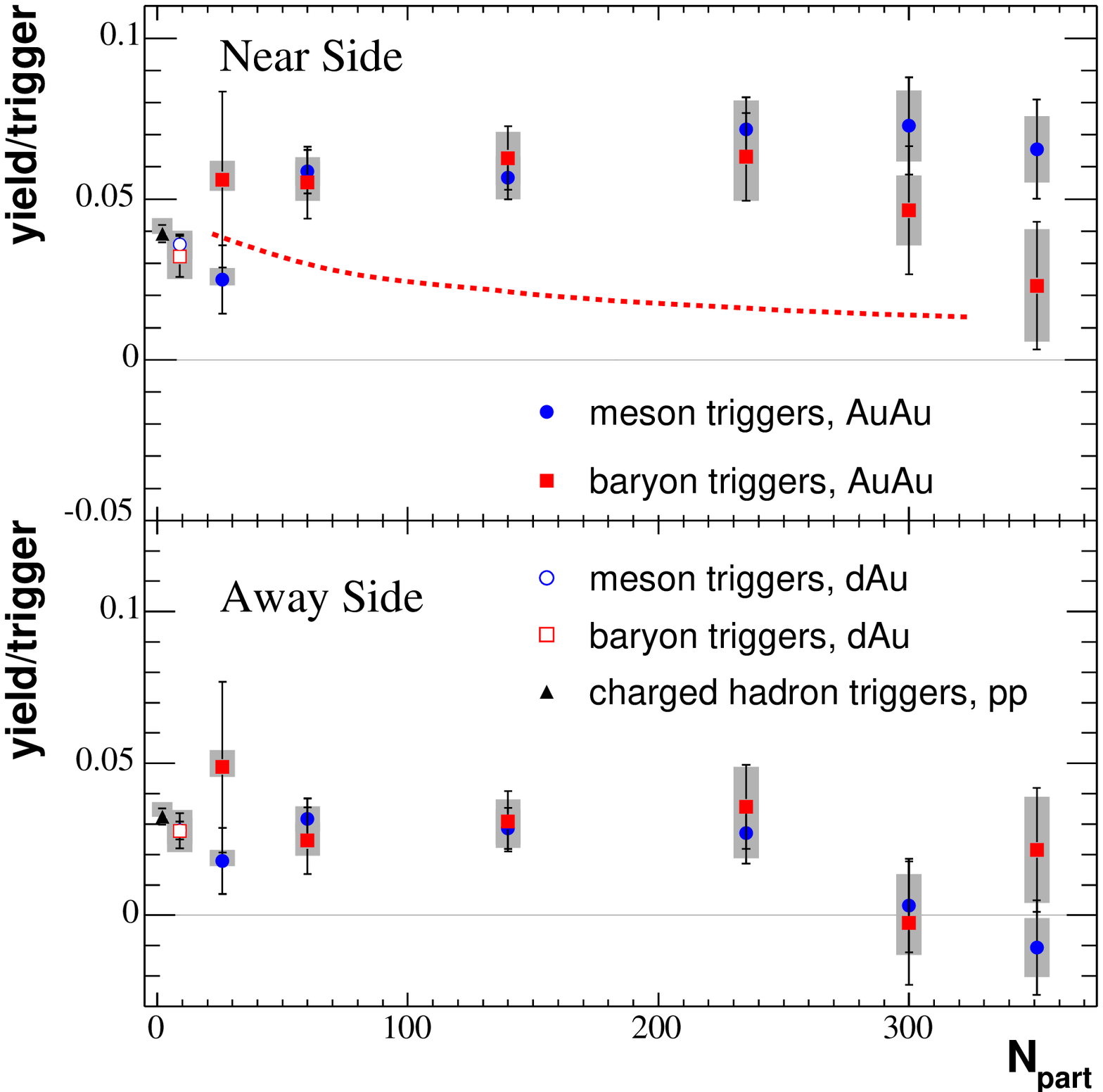}}
\resizebox{0.5\textwidth}{!}{\includegraphics{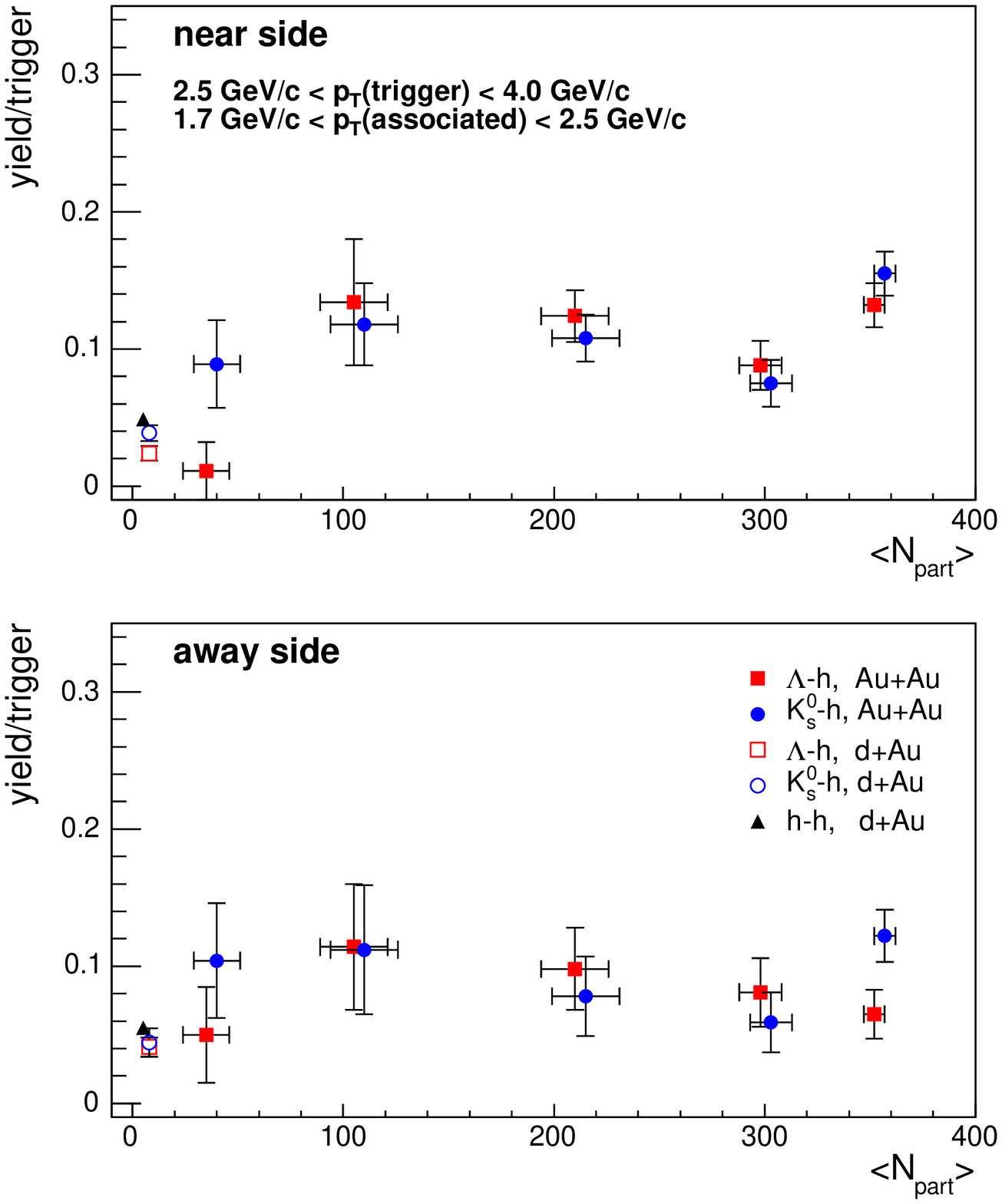}}
\caption{Strength of dihadron azimuthal correlations with identified baryon or meson trigger 
and associated charged hadron, for 200 GeV p+p, d+Au and Au+Au collisions. 
Upper panels show associated yield per trigger for $\Delta\phi\sim0$,
lower panels for $\Delta\phi\sim\pi$.  Left: trigger proton or pion
from PHENIX \cite{Adler:2004zd}. Dashed line is estimated upper limit
of near-side partner yield from thermal recombination. Right: trigger
$\Lambda$ or \kzeros\ from STAR
\cite{Guuo:2005qm,MagestroHardProbes}. The acceptance is
$|\eta|\lt0.35$ for the left panels and $|\eta|\lt0.7$ for the right
panels, with $2.5\lt\pT^{trig}\lt4.0$ GeV/c and
$1.7\lt\pT^{assoc}\lt2.5$ GeV/c for all panels.}
\label{fig:CorrPID}
\end{figure*}

Hadron production via quark coalescence from a thermalized medium will
generate different correlation structure than production via parton
fragmentation, and dihadron correlation measurements therefore provide
additional tests of the recombination picture. The left panels of Fig.
\ref{fig:CorrPID} show measurements by PHENIX of the correlated yield 
measured at small angular separation (``near side'') or back-to-back
(``away side'') relative to triggers identified as pion or proton
\cite{Adler:2004zd}. Significant correlated yield is seen for both
trigger classes, in particular on the near side where the correlation
is stronger for Au+Au collisions than for the d+Au and p+p reference
systems. The dashed line shows results of a calculation in which
thermal recombination is responsible for all excess baryons,
inadequate to account for the observed correlation strength.

The right panels in Fig. \ref{fig:CorrPID} show similar measurements
from STAR for identified $\Lambda$ and \kzeros\ triggers in the same
kinematic window \cite{Guuo:2005qm,MagestroHardProbes}, confirming the
PHENIX observation of similar associated yield for meson and baryon
triggers. (The numerical differences in correlation strength between
the measurements may be attributable to their different $\eta$
acceptances.) The trigger \pT\ interval in these measurements
corresponds to intermediate \pT, where there is a marked difference in
the suppression of meson and baryon inclusive yields in more central
collisions. The near-side associated yield is therefore largely
independent of the degree of inclusive suppression, as would be
expected if these hadrons are generated by jets fragmenting in
vacuum. It is apparent that hadron production at intermediate
\pT\ is dominated neither by energy loss and fragmentation of
hard-scattered partons nor by recombination from an uncorrelated
thermal source. Rather, it exhibits an interplay between parton
fragmentation and hadronization from the bulk medium, with their
relative contributions varying as a function of
\pT\ \cite{Hwa:2004ng,Greco:2003mm}. Dynamical effects may result 
for instance from the excitation of the thermal medium by the partonic
energy loss (``wake-field'' effect)
\cite{Fries:2004hd} or the coupling of medium-induced gluon radiation to bulk longitudinal flow 
\cite{Armesto:2004pt}. 

\begin{figure}
\resizebox{0.5\textwidth}{!}{\includegraphics{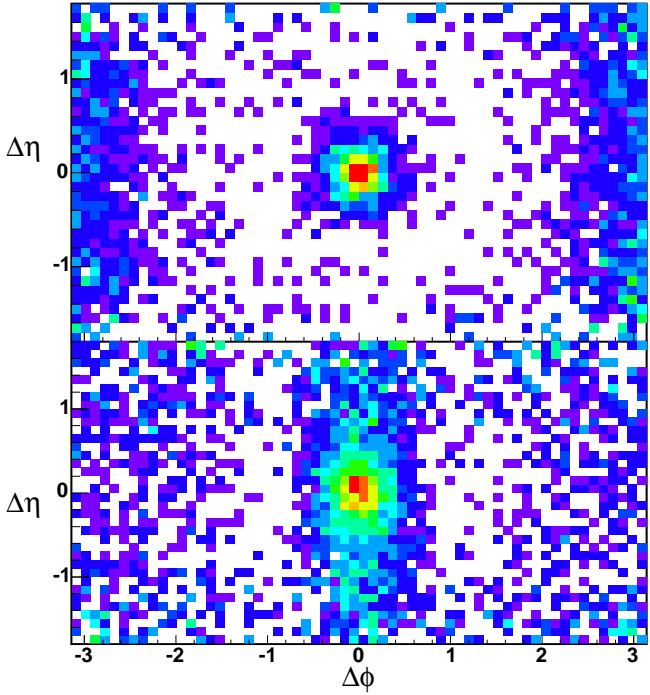}}
\caption{Two-dimensional ($\Delta\eta\times\Delta\phi$) charged dihadron correlation 
functions from minimum bias p+p (upper) and central Au+Au (lower)
collisions at \sqrts=200 GeV, with $3\lt\pT^{trig}\lt6$ GeV/c and
$2\lt\pT^{assoc}\lt\pT^{trig}$
\cite{MagestroHardProbes}.}
\label{fig:CorrMagestro}
\end{figure}

Figure \ref{fig:CorrMagestro} shows new dihadron correlation data
presented by STAR at this conference \cite{MagestroHardProbes}. The
figure shows two dimensional ($\Delta\eta\times\Delta\phi$)
correlation functions for high \pT\ charged dihadrons from p+p (upper)
and central Au+Au (lower) collisions, corrected for finite $\eta$
acceptance. (This is in contrast to Fig. \ref{fig:Corr}, where the
reduced pair acceptance at large longitudinal separation $\Delta\eta$
has not been taken into account, with large $\Delta\eta$ pairs
relatively suppressed as a result.)  Both panels show a jet-like
correlation at small angular separation ($\Delta\eta\sim0$,
$\Delta\phi\sim0$) as well as correlations back-to-back in azimuth
($\Delta\phi\sim\pi$) but broad in $\Delta\eta$ which are presumed to
result from recoil jets and, in the nuclear case, from elliptic
flow. The new aspect is the additional correlation strength for
central Au+Au but not p+p which is short range in azimuth
($\Delta\phi\sim0$) but long range in pseudorapidity
($\Delta\eta\sim$large), and which is stronger than the recoil
correlation at $\Delta\phi\sim\pi$. Preliminary analysis suggests that
this long-range component is effectively uniform within the STAR
acceptance $|\eta|\lt1$ and is distinct from the jet-like peak,
perhaps indicating an independent underlying mechanism. Near-side
long-range correlations for soft hadrons with \pT\lt2 GeV/c were first
noted in \cite{Adams:2004pa}. The new analysis extends the
observations to the region where parton fragmentation may play a
signficant role. The width of the short range jet-like correlation is
similar in central Au+Au and p+p collisions for $\pT^{trig}\gt6$ GeV/c
but exhibits medium-induced broadening a lower $\pT^{trig}$, while the
correlated yield of the jet-like peak is independent of centrality
\cite{MagestroHardProbes}, consistent with Fig. \ref{fig:Corr}. 

The physics underlying these near-side effects is not yet clear. The
long-range correlation may perhaps be due to the combined effect of
the geometric trigger bias and strong transverse radial flow of the
bulk
\cite{Voloshin:2003ud}, or coupling of induced radiation to
longitudinal flow \cite{Armesto:2004pt}. Systematic studies of the
\pT-dependence of the correlation will help resolve these issues. 
Evidence for coupling of induced radiation to bulk longitudinal flow
would indicate non-negligible energy loss, calling into question the
simple picture of geometric (surface) bias of the leading particle
trigger used to interpret other high \pT\ data.

%------------------------------------------------------------------------------
\section{The Future: RHIC II and the LHC}
\label{Future}

The above results are mainly from the 2002 Au+Au run at RHIC, which
delivered integrated luminosity of about 250 $\mu$b$^{-1}$. The much
higher statistics 2004 dataset for Au+Au at 200 GeV now being analysed
has $\sim3.7$ nb$^{-1}$, corresponding to about twice the RHIC design
luminosity \cite{HarrisHardProbes}. Significant RHIC detector upgrades
over the next five years are aimed at improved vertexing for charm
measurements, better particle identification, and low-mass dilepton
measurements
\cite{HarrisHardProbes}; and expanded forward coverage to 
address low $x$ physics \cite{BlandHardProbes}. The five year run plan
calls for collisions of lighter ions, extensive runs with lower energy
Au+Au, polarized p+p and d+Au, and another long 200 GeV Au+Au run once
major detector upgrades are in place. In the longer term, a
significant accelerator upgrade based on electron cooling is planned
(RHIC II) which will give annual integrated luminosity for 200 GeV
Au+Au of $\sim80-90$ nb$^{-1}$. Additional upgrades to the existing
large detectors and an entirely new detector are being considered to
exploit the increased luminosity
\cite{HarrisHardProbes} which will extend jet-related
measurements to much higher \pT, far into the perturbative regime. The
broad scientific case for RHIC II is under discussion \cite{RHICtwo},
but for hard probes the increased capabilities will enable much more
robust $\gamma$+jet measurements and detailed study of the charm and,
in particular, beauty sectors.

First p+p collisions at the LHC are currently expected in 2007, with
first Pb+Pb collisons at 5.5 TeV in 2008. Jet rates at the LHC are
large
\cite{Accardi:2003gp}. The nominal 
annual integrated luminosity for 5.5 TeV Pb+Pb collisions is 0.5
nb$^{-1}$, yielding $\sim1.4\times10^6$ jets with $\ET\gt150$ GeV and
$\sim3.7\times10^5$ jets with $\ET\gt200$ GeV within acceptance
$|\eta|\lt2.5$. The broad kinematic and statistical reach of jet
production at the LHC enables study over a very broad range of
fragmentation $z$ of the softening and transverse heating of jets due
to interactions in the medium
\cite{Salgado:2003rv,Vitev:2005yg}. Partonic energy loss is expected
to vary at most logarithmically with jet energy, and the wide range of
jet \ET\ will provide sufficient lever arm for a stringent test of
energy loss theory. Nearly complete reconstruction of jets may be
possible in LHC heavy ion collisions. Complete jet reconstruction
would be free of the geometric and fragmentation biases intrinsic to
leading particle triggers and would be sensitive to the full
$A^2$-scaled jet cross section, giving access to the full spectrum of
energy loss in the medium.

While the identification of hard jets at RHIC II and the LHC will be
straightforward, full jet reconstruction with good energy resolution
must contend with the complexity of the underlying background in heavy
ion collisions. The average high \ET\ jet at the Tevatron has about
80\% of its energy and multiplicity within a cone of radius $R=0.3$
\cite{Affolder:2001xt}, while a Pb+Pb event with
$\dNchdeta\sim4000$ at the LHC will deposit $\sim100$ GeV of
background energy in the same cone. Current studies indicate that
$R\sim0.3-0.4$ gives optimal signal/background for jet reconstruction
in central Pb+Pb collisions, with correspondingly limited energy
resolution
\cite{MorschHardProbes}. There are large uncertainties 
in the modelling of fragmentation and background fluctuations used for
these estimates, however, and the first LHC data will tell us much
more about jet triggering and reconstruction in nuclear collisions at
5.5 TeV.

The $\gamma$+jet final state is expected to provide the most
controlled measurement of jet quenching, since the photon does not
interact with the colored medium and its \pT\ balances that of the
recoiling jet
\cite{Wang:1996pe}. This channel is likewise free of the 
geometric bias of leading particle triggers. PHENIX has measured
inclusive direct photons in the heavy ion environment at RHIC, helped
by the hadron suppression which reduces background
\cite{Frantz:2004gg,ReygersHardProbes}. The $\gamma$+jet coincidence measurement 
is statistically more demanding, however, and a robust measurement
extending well above $\pT^\gamma\sim10$ GeV/c will require the RHIC II
luminosity upgrade \cite{HarrisHardProbes}. The higher
\pT\ is important to provide dynamic range in fragmentation $z$ and to
reduce the fragmentation photon contribution
\cite{ReygersHardProbes}. Hard cross sections at the LHC are substantially larger and 
statistically significant $\gamma$+jet measurements in Pb+Pb may
extend to $\pT^\gamma\sim40$ GeV. Fragmentation photons are expected
to dominate the cross section in this region, however, and their
contribution will have to be disentangled \cite{Arleo:2004xj}. Z+jet
provides a very clean but statistically more limited alternative
\cite{WyslouchHardProbes}. 

A crucial test of the partonic energy loss picture is the variation of
energy loss with parton species. Generically, gluons should lose more
energy than quarks due to their larger color charge. It has
also been proposed that heavy quarks should experience smaller energy
loss than light quarks due to the mass-dependent suppression of
forward radiation (``dead cone effect'' 
\cite{Dokshitzer:2001zm,Armesto:2003jh,Zhang:2003wk,Djordjevic:2003zk}). Isolation 
of specifically quark or gluon jet effects at RHIC has proven difficult
thus far, but forward triggering at moderately high \pT\ may provide a
controlled bias towards gluon jets
\cite{BlandHardProbes}. Measurements of heavy quark production at RHIC
are developing rapidly \cite{ScomparinHardProbes}. It was recently
shown that the suppression of D and B mesons relative to light mesons
is influenced not only by the mass-dependent dead cone effect but also
by color-charge dependence, especially at the LHC where gluon jets
dominate light meson production \cite{Armesto:2005iq}. Measurements of
D and B production over a broad \pT\ range at RHIC and the LHC will
map out these effects in detail.

%----------------------------------------------------------------------------
\section{Summary}
\label{Summary}

Medium-induced modification of jet structure provides a powerful tool
for the study of QCD matter created in high energy nuclear
collisions. Jet modification effects at RHIC are large and
statistically robust, enabling their detailed
characterization. Perturbative calculations incorporating partonic
energy loss via gluon bremsstrahlung agree well with the measurements,
provided the medium has gluon density much greater than that of cold
nuclear matter. The intermediate region $\sim2\lt\pT\lt5$ GeV/c
exhibits an apparent interplay between fragmentation of hard partons
and flow of the bulk medium that may provide a window to the partonic
processes that drive the system to equilibrium. However, important
aspects of energy loss theory have not yet been tested. Measurements
in the near future at RHIC and the LHC will push the kinematic
boundaries of jet studies significantly outward and provide strong
constraints on the physics underlying partonic energy loss. It remains
to be seen whether the intermediate
\pT\ region at the LHC will exhibit similar interplay as at RHIC between
perturbative and non-perturbative processes, but it is reasonable to
expect that this region will remain interesting even for fragments of
the highest energy jets at RHIC II and the LHC.
%----------------------------------------------------------------------------
\section{Acknowledgements}
\label{Acknowledgements}

I thank Urs Wiedemann for many helpful discussions.
%---------------------------------------------------------------------------------

%\bibliographystyle{unsrt}
\bibliographystyle{h-physrev}
\bibliography{HardProbes.bib}

\begin{thebibliography}{10}

\bibitem{Adcox:2004mh}
PHENIX, K.~Adcox {\em et~al.},
\newblock nucl-ex/0410003.

\bibitem{Back:2004je}
B.~B. Back {\em et~al.},
\newblock nucl-ex/0410022.

\bibitem{Arsene:2004fa}
BRAHMS, I.~Arsene {\em et~al.},
\newblock nucl-ex/0410020.

\bibitem{adams:2005dq}
STAR, J.~Adams,
\newblock nucl-ex/0501009.

\bibitem{BlandHardProbes}
L.~Bland {\em et~al.},
\newblock hep-ex/0502040.

\bibitem{Frantz:2004gg}
PHENIX, J.~Frantz,
\newblock J. Phys. {\bf G30}, S1003 (2004), nucl-ex/0404006.

\bibitem{BueschingHardProbes}
H.~Buesching,
\newblock these proceedings.

\bibitem{Adler:2003pb}
PHENIX, S.~S. Adler {\em et~al.},
\newblock Phys. Rev. Lett. {\bf 91}, 241803 (2003), hep-ex/0304038.

\bibitem{Baier:2000mf}
R.~Baier, D.~Schiff, and B.~G. Zakharov,
\newblock Ann. Rev. Nucl. Part. Sci. {\bf 50}, 37 (2000), hep-ph/0002198.

\bibitem{Gyulassy:2003mc}
M.~Gyulassy, I.~Vitev, X.-N. Wang, and B.-W. Zhang,
\newblock (2003), nucl-th/0302077.

\bibitem{Kovner:2003zj}
A.~Kovner and U.~A. Wiedemann,
\newblock hep-ph/0304151.

\bibitem{Baier:2002tc}
R.~Baier,
\newblock Nucl. Phys. {\bf A715}, 209 (2003), hep-ph/0209038.

\bibitem{Eskola:2004cr}
K.~J. Eskola, H.~Honkanen, C.~A. Salgado, and U.~A. Wiedemann,
\newblock Nucl. Phys. {\bf A747}, 511 (2005), hep-ph/0406319.

\bibitem{Salgado:2003gb}
C.~A. Salgado and U.~A. Wiedemann,
\newblock Phys. Rev. {\bf D68}, 014008 (2003), hep-ph/0302184.

\bibitem{WangHardProbes}
X.-N. Wang,
\newblock these proceedings.

\bibitem{WiedemannHardProbes}
U.~A. Wiedemann,
\newblock hep-ph/0503119.

\bibitem{Adler:2002tq}
STAR, C.~Adler {\em et~al.},
\newblock Phys. Rev. Lett. {\bf 90}, 082302 (2003), nucl-ex/0210033.

\bibitem{Drees:2003zh}
A.~Drees, H.~Feng, and J.~Jia,
\newblock (2003), nucl-th/0310044.

\bibitem{Dainese:2004te}
A.~Dainese, C.~Loizides, and G.~Paic,
\newblock Eur. Phys. J. {\bf C38}, 461 (2005), hep-ph/0406201.

\bibitem{d'Enterria:2004fm}
D.~d'Enterria,
\newblock nucl-ex/0406012.

\bibitem{ColeHardProbes}
B.~Cole,
\newblock these proceedings.

\bibitem{Vitev:2002pf}
I.~Vitev and M.~Gyulassy,
\newblock Phys. Rev. Lett. {\bf 89}, 252301 (2002), hep-ph/0209161.

\bibitem{Wang:2001cs}
E.~Wang and X.-N. Wang,
\newblock Phys. Rev. Lett. {\bf 87}, 142301 (2001), nucl-th/0106043.

\bibitem{dEnterriaHardProbes}
D.~d'Enterria,
\newblock these proceedings.

\bibitem{GagliardiHardProbes}
C.~Gagliardi,
\newblock these proceedings.

\bibitem{Back:2004ra}
PHOBOS, B.~B. Back {\em et~al.},
\newblock nucl-ex/0405003.

\bibitem{Vitev:2004gn}
I.~Vitev,
\newblock nucl-th/0404052.

\bibitem{Adil:2004cn}
A.~Adil and M.~Gyulassy,
\newblock Phys. Lett. {\bf B602}, 52 (2004), nucl-th/0405036.

\bibitem{Wang:2004yv}
X.-N. Wang,
\newblock Phys. Rev. {\bf C70}, 031901 (2004), nucl-th/0405029.

\bibitem{Adams:2003im}
STAR, J.~Adams {\em et~al.},
\newblock Phys. Rev. Lett. {\bf 91}, 072304 (2003), nucl-ex/0306024.

\bibitem{Adams:2005ph}
STAR, J.~Adams {\em et~al.},
\newblock nucl-ex/0501016.

\bibitem{Borghini:2000cm}
N.~Borghini, P.~M. Dinh, and J.-Y. Ollitrault,
\newblock Phys. Rev. {\bf C62}, 034902 (2000), nucl-th/0004026.

\bibitem{Casalderrey-Solana:2004qm}
J.~Casalderrey-Solana, E.~V. Shuryak, and D.~Teaney,
\newblock hep-ph/0411315.

\bibitem{Adams:2004wz}
STAR, J.~Adams {\em et~al.},
\newblock Phys. Rev. Lett. {\bf 93}, 252301 (2004), nucl-ex/0407007.

\bibitem{Adler:2003kg}
PHENIX, S.~S. Adler {\em et~al.},
\newblock Phys. Rev. Lett. {\bf 91}, 172301 (2003), nucl-ex/0305036.

\bibitem{Adams:2003am}
STAR, J.~Adams {\em et~al.},
\newblock Phys. Rev. Lett. {\bf 92}, 052302 (2004), nucl-ex/0306007.

\bibitem{VelkovskaHardProbes}
J.~Velkovska,
\newblock these proceedings.

\bibitem{Voloshin:2002wa}
S.~A. Voloshin,
\newblock Nucl. Phys. {\bf A715}, 379 (2003), nucl-ex/0210014.

\bibitem{Fries:2003vb}
R.~J. Fries, B.~Muller, C.~Nonaka, and S.~A. Bass,
\newblock Phys. Rev. Lett. {\bf 90}, 202303 (2003), nucl-th/0301087.

\bibitem{Hwa:2004ng}
R.~C. Hwa and C.~B. Yang,
\newblock Phys. Rev. {\bf C70}, 024905 (2004), nucl-th/0401001.

\bibitem{Greco:2003xt}
V.~Greco, C.~M. Ko, and P.~Levai,
\newblock Phys. Rev. Lett. {\bf 90}, 202302 (2003), nucl-th/0301093.

\bibitem{Adler:2004zd}
PHENIX, S.~S. Adler {\em et~al.},
\newblock nucl-ex/0408007.

\bibitem{Guuo:2005qm}
Y.~Guo,
\newblock nucl-ex/0502015.

\bibitem{MagestroHardProbes}
D.~Magestro,
\newblock these proceedings.

\bibitem{Greco:2003mm}
V.~Greco, C.~M. Ko, and P.~Levai,
\newblock Phys. Rev. {\bf C68}, 034904 (2003), nucl-th/0305024.

\bibitem{Fries:2004hd}
R.~J. Fries, S.~A. Bass, and B.~Muller,
\newblock nucl-th/0407102.

\bibitem{Armesto:2004pt}
N.~Armesto, C.~A. Salgado, and U.~A. Wiedemann,
\newblock Phys. Rev. Lett. {\bf 93}, 242301 (2004), hep-ph/0405301.

\bibitem{Adams:2004pa}
STAR, J.~Adams {\em et~al.},
\newblock nucl-ex/0411003.

\bibitem{Voloshin:2003ud}
S.~A. Voloshin,
\newblock nucl-th/0312065.

\bibitem{HarrisHardProbes}
J.~W. Harris,
\newblock nucl-ex/0503014.

\bibitem{RHICtwo}
http://www.bnl.gov/physics/rhicIIscience/default.asp.

\bibitem{Accardi:2003gp}
A.~Accardi {\em et~al.},
\newblock hep-ph/0310274.

\bibitem{Salgado:2003rv}
C.~A. Salgado and U.~A. Wiedemann,
\newblock Phys. Rev. Lett. {\bf 93}, 042301 (2004), hep-ph/0310079.

\bibitem{Vitev:2005yg}
I.~Vitev,
\newblock hep-ph/0501255.

\bibitem{Affolder:2001xt}
CDF, T.~Affolder {\em et~al.},
\newblock Phys. Rev. {\bf D65}, 092002 (2002).

\bibitem{MorschHardProbes}
A.~Morsch,
\newblock these proceedings.

\bibitem{Wang:1996pe}
X.-N. Wang and Z.~Huang,
\newblock Phys. Rev. {\bf C55}, 3047 (1997), hep-ph/9701227.

\bibitem{ReygersHardProbes}
PHENIX, K.~Reygers,
\newblock nucl-ex/0502018.

\bibitem{Arleo:2004xj}
F.~Arleo, P.~Aurenche, Z.~Belghobsi, and J.-P. Guillet,
\newblock JHEP {\bf 11}, 009 (2004), hep-ph/0410088.

\bibitem{WyslouchHardProbes}
B.~Wyslouch,
\newblock these proceedings.

\bibitem{Dokshitzer:2001zm}
Y.~L. Dokshitzer and D.~E. Kharzeev,
\newblock Phys. Lett. {\bf B519}, 199 (2001), hep-ph/0106202.

\bibitem{Armesto:2003jh}
N.~Armesto, C.~A. Salgado, and U.~A. Wiedemann,
\newblock Phys. Rev. {\bf D69}, 114003 (2004), hep-ph/0312106.

\bibitem{Zhang:2003wk}
B.-W. Zhang, E.~Wang, and X.-N. Wang,
\newblock Phys. Rev. Lett. {\bf 93}, 072301 (2004), nucl-th/0309040.

\bibitem{Djordjevic:2003zk}
M.~Djordjevic and M.~Gyulassy,
\newblock Nucl. Phys. {\bf A733}, 265 (2004), nucl-th/0310076.

\bibitem{ScomparinHardProbes}
E.~Scomparin,
\newblock these proceedings.

\bibitem{Armesto:2005iq}
N.~Armesto, A.~Dainese, C.~A. Salgado, and U.~A. Wiedemann,
\newblock (2005), hep-ph/0501225.

\end{thebibliography}

\end{document}